\documentclass[useAMS,usegraphicx,usenatbib]{mn2e}
\usepackage{times}
\usepackage{url} 

\title[X-ray spectra of tetrahedral nanodiamonds]{ Tetrahedral hydrocarbon nanoparticles in space: X-ray spectra}
\author[G. Bilalbegovi\' c, A. Maksimovi\' c,  L. A. Valencic]
{G. Bilalbegovi\' c$^{1}$,
A. Maksimovi\' c$^{2}$,
L. A. Valencic$^{3,4}$
\\
$^{1}$Department of Physics, Faculty of Science, University of Zagreb, Bijeni\v cka cesta 32, 10000 Zagreb, Croatia\\
$^{2}$Center of Excellence for Advanced Materials and Sensing Devices, Rudjer Bo\v skovi\' c Institute, Bijeni\v  cka cesta 54, 10000 Zagreb, Croatia\\
$^{3}$NASA Goddard Space Flight Center, Greenbelt, MD 20771, USA\\
$^{4}$The Johns Hopkins University, Department of Physics \& Astronomy, 366 Bloomberg Center, 3400 N. Charles St., Baltimore, MD 21218, USA
}

\begin{document}

\date{\today}
%\date{Accepted . Received ; in original form} 

\pagerange{\pageref{firstpage}--\pageref{lastpage}}
\pubyear{2018} \volume{000}

\maketitle 
\label{firstpage}

\begin{abstract}

It has been proposed, or confirmed, that diamond nanoparticles exist in various environments in space: close to active galactic nuclei, in the vicinity of supernovae and pulsars,  in the interior of several planets in the Solar system, in carbon planets and other exoplanets, carbon-rich stars, meteorites, in X-ray active Herbig Ae/Be stars, and in the interstellar medium. Using density functional theory methods we calculate the carbon K-edge X-ray absorption spectrum of  two large tetrahedral nanodiamonds: C$_{26}$H$_{32}$ and C$_{51}$H$_{52}$. We also study and test our methods on the astrophysical molecule CH$_4$, the smallest C-H tetrahedral structure. A possible detection of nanodiamonds  from X-ray spectra by future  telescopes, such as  the project {\it Arcus}, is proposed. Simulated spectra of the diffuse interstellar medium using Cyg X-2 as a source show that nanodiamonds studied in this work can be detected by {\it Arcus}, a high resolution X-ray spectrometer mission selected by NASA for a Phase A concept study.

\end{abstract}


\section{Introduction}
\label{intro}

Nanoparticles of diamonds (also known as nanodiamonds, diamondoids, diamondoid molecules, diamond clusters) are structures important in various branches of science \citep{Galli2010,Clay2015}.  
\cite{Saslaw1969} suggested a long time ago that diamond structures could play an important role in astrophysical phenomena. 
An earlier work about diamonds in the astrophysical context exists \citep{Wentorf1961}.
The authors compared ''the diamonds of men and the diamonds of nature'' to shed light on  their formation. 
However, the question of how diamonds form in space is still not completely solved.

\subsection{Meteorites and rocks on Earth}
Nanodiamonds were found at several locations in the Younger Dryas boundary sediments \citep{Kennett2009}. It was proposed that these nanoparticles formed during impacts of comets or asteroids. 
Nanoparticles of diamonds are the most abundant form of carbon in carbonaceous chondrite meteorites
\citep{Lewis1987,Lewis1989}. It is accepted today that these meteoritic nanodiamonds were formed in the Solar system by processes similar to chemical vapor deposition \citep{Daulton1996,Dai2002}.

\subsection{Supernovae, the interstellar medium, pulsars, carbon planets and stars}
 It was proposed that diamond nanoparticles form in dust processing by strong shock waves in the vicinity of a supernovae \citep{Tielens1987}.  \cite{Tielens1987} calculated  that approximately 5\% of the carbon in the ISM exists in the form of (5-100) \AA\, nanodiamonds. 

Nanodiamonds were suggested  as carriers of extended red emission \citep{Duley1988,Chang2006}.
\cite{Rai2012} found that nanodiamonds are good candidates for a source of the observed extinction along anomalous sightlines.

It was estimated that the binary object of the pulsar J1719-1438 has the minimum density of 23 g/cm$^3$ \citep{Bailes2011}. The chemical composition, dimensions, and pressures of J1719-1438 b
indicate that it consists of diamonds.
\cite{Kuchner2005} proposed the existence of such carbon planets that contain more carbon than oxygen  and suggested that substantial amount of diamond should exists there. The carbon-rich composition and the existence of diamonds  has been predicted for the pulsar planets \citep{Margalit2017}.
\cite{Helling2017} presented a kinetic model of cloud formation around carbon-rich exoplanets and brown dwarfs. They found that inner parts of carbon-rich clouds could be made of diamonds.
It was also proposed that diamonds form in the envelopes of carbon-rich stars, 
such as HR 4049 that is at the post-AGB phase of evolution  \citep{Geballe1989,VanKerckhoven2002}.

\subsection{Active galactic nuclei} 
A dust model consisting of nanodiamonds was one of approaches used to explain the big blue bump
 in the spectral energy distribution of active galactic nuclei  (AGNs) \citep{HaroCorzo2007}.
AGNs were also suggested as nanodiamond factories \citep{Rouan2004,Gratadour2006}.
Nanodiamonds efficiently form in UV and shocks \citep{Jones2000}. 
They could survive X-ray and UV radiation under conditions where silicate dust grains are destroyed.

\subsection{Herbig Ae/Be stars}
The infrared (IR) features at 3.43 and 3.53 $\mu$m  in the spectra of the Herbig Ae/Be (HAeBe) stars Elias 1 (V892 Tau) and HD 97048  (CU Chamaeleontis)  were explained by the C-H stretching vibrational modes in hydrogen-saturated diamonds present in protoplanetary disks \citep{Guillois1999,Jones2000,VanKerckhoven2002,Habart2004,Topalovic2006,Pirali2007,Goto2009}.  
Majority of stars in the HAeBe class do not have spectral features at 3.43 and 3.53 $\mu$m \citep{Acke2006}.  
Elias 1 and HD 97048 were observed as sources of  X-ray emission  by several telescopes, for example in
the ROSAT survey \citep{Zinnecker1994}, by {\it ASCA}  \citep{Hamaguchi2005}, as well as by XMM-{\it Newton} and {\it Chandra} \citep{Giardino2004,Skinner2004,Franciosini2007}. 

From IR observations it is known that 85\% of HAeBe stars are binaries \citep{Pirzkal1997}.
It was proposed that nanodiamonds form in conditions where a protoplanetary disk exists around a hot star, with a companion emitting X-rays \citep{Goto2009}.  
Elias 1 is a triple star  system  in the Taurus-Auriga complex \citep{Skinner1993,Leinert1997,Giardino2004,Smith2005,Franciosini2007}.  
IR observations indicated that Elias 1 has a circumstellar disk of $\sim$0.1 M$_{\odot}$ \citep{DiFrancesco1997}. 
The first companion Elias 1 NE is in the T Tauri class, and it is located 4'' north-east of Elias 1. The second companion was discovered by \cite{Smith2005}. This convective star at 0.05'', with mass of 1.5-2 M$_{\odot}$,  was suggested as the source of X-rays. 
\cite{Giardino2004} found a strong X-ray flare in the triple Elias system.
However, the presence of companions is not the only possible explanation of X-ray emission of HAeBe stars. For example, stellar winds and star-disk magnetospheres have been studied \citep{Stelzer2006}.

HD 97048 is a young star in the constellation Chamaeleon with a pronounced dusty disk \citep{Habart2004,Doering2007}.
The dynamo model was proposed to explain X-ray emission in HD 97048 \citep{Tout1995,Skinner2004}. 
A protoplanetary disk with multiple rings and gaps around HD 97048 was recently observed \citep{Ginski2016,Walsh2016,vanderPlas2017}.

\subsection{Methane}
Methane molecules are present  in Earth's atmosphere where they could interact with scattered solar and auroral
X-rays.  CH$_4$ was also observed in several extraterrestrial sources \citep{Mousis2015}. For example, the presence of methane  was confirmed by Cassini spectrometers on the Saturn`s largest moon Titan \citep{Niemann2005}, as well as
in  the Enceladus plume \citep{Waite2006}.  X-ray emission from Saturn was observed by XMM-Newton and Chandra \citep{Branduardi2010}.
CH$_4$  was detected in comets \citep{Gibb2003} and several comets were found to emit X-rays \citep{Lisse1996,Dennerl1997,Lisse2001}. 
Methane was also found in extrasolar planets \citep{Swain2008}, the atmospheres of M-type stars \citep{Jorgensen1996}, and the interstellar medium \citep{Lacy1991}.   
\cite{Kraus2017} recently performed the experiment on formation of diamonds in laser-compressed hydrocarbons. Conditions of this experiment simulated interiors of Neptune, Uranus and exoplanets. It is known that methane is very abundant in the atmospheres of Neptune, Uranus and some exoplanets. Methane transforms into a mixture of hydrocarbon polymers close to the surface of Uranus and Neptune. A phase separation into diamond and hydrogen is possible 
in deeper layers  \citep{Kraus2017}.
A transformation of CH$_4$ into diamond also occurs in the interior of icy super-earth mass planets \citep{Levi2017}.

\subsection{Nanodiamonds}
 Surface atoms of nanodiamonds have unpaired bonds that are saturated with H atoms in a hydrogen rich atmosphere. The smallest nanodiamonds are adamantane (C$_{10}$H$_{16}$), diamantane (C$_{14}$H$_{20}$) and triamantane (C$_{18}$H$_{24}$). 
Isomers for higher nanodiamonds  (starting from tetramantane C$_{22}$H$_{28}$) exist, and only some of them are tetrahedral. 
\cite{Oomens2006} studied IR spectra for several smaller nanodiamonds, from adamantane to hexamantane, and compared with astrophysical observations of V892 Tau and HD 97048. 
In addition, \cite{Pirali2007} extrapolated intensities of  IR spectra
of smaller nanodiamonds they studied. They found a good agreement between spectral intensities
of larger tetrahedral diamondoids C$_{51}$H$_{52}$ and C$_{87}$H$_{76}$  and observed ones for Elias 1 and HD 97048.
X-ray spectra of small nanodiamonds (up to hexamantane) were measured  \citep{Willey2005} but, to the best of our knowledge, similar studies on larger nanodiamonds do not exist. It is difficult to prepare large nanodiamonds on Earth by organic synthesis, or to extract them from petroleum.
The largest tetrahedral nanodiamond discussed by \cite{Pirali2007}, C$_{87}$H$_{76}$,
is too large for available computational resources. Therefore,
we calculate the K-edge X-ray spectrum of three tetrahedral hydrocarbon structures:
CH$_4$,   the [1(2,3)4] pentamantane isomer of  C$_{26}$H$_{32}$, and C$_{51}$H$_{52}$.
The nomenclature of isomers for diamondoids was introduced by \cite{Balaban1978}.

Studies of X-ray spectra for materials in the astrophysical context are rather new, and till now done mainly for iron and silicate dust compounds using experimental synchrotron based methods \citep{Lee2005,Lee2009,Zeegers2017,Rogantini2018}. Calculations of scattering and absorption of X-rays (based on  realistic dielectric functions) by carbonaceous and silicate dust grains  are also available \citep{Draine2003}.
In this work density functional theory (DFT) is applied to calculate the X-ray absorption spectrum  of tetrahedral hydrocarbon nanoparticles.

\section{Computational methods}
\label{methods}

DFT methods \citep{Jones2015}, as implemented in the GPAW  and ASE software suites \citep{Enkovaara2010,Larsen2017}, are used. 
DFT is a technique for modeling properties of quantum many-body systems by numerically calculating their electron density. We use DFT to study three tetrahedral hydrocarbon finite structures.
The PBE generalized gradient approximation (GGA) \citep{Perdew1996} and the projector augmented wave method 
(PAW) \citep{Blochl1994,Mortensen2005}
are applied. The PAW method enables the reconstruction of the all-electron wave function and better modeling of  X-ray spectra \citep{Ljungberg2011}. The pseudopotential of the absorbing atom includes core-hole effects. 
The initial C 1s occupation is set to 0.5,  in accordance with the transition potential approximation \citep{Triguero1998}.
We have carried out half and full core calculations 
for all carbon atoms in all structures we study. The size of a computational box and the number of unoccupied states are investigated. Because of large sizes of nanodiamonds (in a comparison with simple molecules),  we use the Haydock recursion method \citep{Ljungberg2011,Haydock1972} to calculate spectra and cover completely the post-edge region. 
The absolute energy scale is determined from the separate Delta Kohn-Sham calculations \citep{Ljungberg2011}. 
Discrete transitions are broadened with Gaussians of the 0.5 eV width.

X-ray spectra we calculate show absorption of a sample as a function of energy. 
The energies of absorption edges are unique for each material sample and depend on its chemical composition, structure, electronic and vibrational properties. 
Spectral peak intensities in DFT  studies of X-ray absorption are derived from the oscillator strengths
which are dimensionless quantities \citep{StohrBook}.  
As in other studies, we present the X-ray absorption intensity in arbitrary units.   Although the exact values of dimensionless oscillator strengths are not important for comparison with experiments on Earth and observations of X-ray telescopes, it is appropriate to further develop theoretical methods to get the best values for these quantities. It is possible to compare the ratio of intensities for calculated oscillator strengths
with the ratio of measured peaks. 
As an example, for hexamantane the ratio of the measured pre-edge peak intensity  and the intensity of the K-edge is 0.8 \citep{Willey2005}.  
The same ratio we calculate for oscillator strengths is 0.4.  This disagreement shows
that DFT calculations are not quantitative enough to obtain oscillator strengths.  However, energies of 
K and other edges are used to compare results with experiments in applications of X-ray absorption spectra.
Good values for K-edges and overall agreement of our results and experimental spectra for methane \citep{Hitchcock1994} and smaller nanodiamonds 
\citep{Willey2005}, as well as results for other systems calculated using the same methods \citep{Triguero1998,Ljungberg2011,Leetmaa2010,Susi2015,Sainio2016,LaRue2017} 
show that DFT methods we use are satisfactory.
Other methods  to calculate X-ray absorption spectra exist. Some of them are more rigorous than DFT, such as the many-body perturbation theory methods based on the Bethe–-Salpeter equation \citep{Shirley1998,Rehr2000,Gilmore2015}. However, such methods are  substantially more computationally demanding, especially for finite large structures we study in this work.

\section{Results and Discussion}
\label{results}

\subsection{X-ray spectra}

The calculated carbon K-edge spectrum of methane is compared with the experimental result in Fig. 1. 
The corrected experimental  electron energy-loss spectrum (EELS) from the database \citep{Hitchcock1994} is used. 
The calculated (288.15 eV) and measured (288.12 eV) main peak differ by 0.01\%. Methods we use do not produce a small 
pre-edge feature present in the EELS data. This feature is also absent in the spectrum of the gas phase CH$_4$ obtained by 
STOBE-DEMON DFT code \citep{Ostrom2006}.  Such a small pre-edge peak appears in the high-resolution experimental spectra and the highest level {\it ab initio} theories because of a specific vibrational coupling effect on a dipole-forbidden transition    
\citep{Schirmer1993}.

The structures of C$_{26}$H$_{32}$ and C$_{51}$H$_{52}$  are shown in Fig. 2. 
Their X-ray spectra  are presented in Figs. 3 and 4.
The main broad peaks are positioned at 292 eV.  Several smaller features  appear from
below of 287 eV  up to the main peak.
Pre-edge structures are more emphasized in the smaller nanodiamond C$_{26}$H$_{32}$.
Similar pre-edge structures exist in experimental X-ray spectra of small nanodiamonds, from adamantane to hexamantane  \citep{Willey2005}. These pre-edge features  are contributions of  the C-H and C-H$_2$ bonds
at surfaces of nanodiamonds.

The internal structures of C$_{26}$H$_{32}$ and C$_{51}$H$_{52}$ resemble one of the bulk diamond (see Fig. 2).
Features of the bulk diamond spectra for both C$_{26}$H$_{32}$ and C$_{51}$H$_{52}$ exist 
slightly below and at 305 eV (see Figs. 3 and 4). These bulk features are more pronounced for larger  C$_{51}$H$_{52}$. However, spectra of large
nanodiamonds C$_{26}$H$_{32}$ and C$_{51}$H$_{52}$  are still rather different from one of the bulk diamond. Our Figs. 3 and 4 are compared with measured \citep{Ma1992} and calculated by the same GPAW methods \citep{Ljungberg2011} spectra of the bulk diamond.
A different type of X-ray spectra of nanodiamonds we study and of the bulk diamond is in contrast to their optical spectra. It was measured that the optical response of the tetrahedral C$_{26}$H$_{32}$  isomer (shown in Fig. 2(a)) is similar to that of bulk diamond \citep{Landt2009}. 

\begin{figure}
\centering 
\includegraphics[scale=0.4]{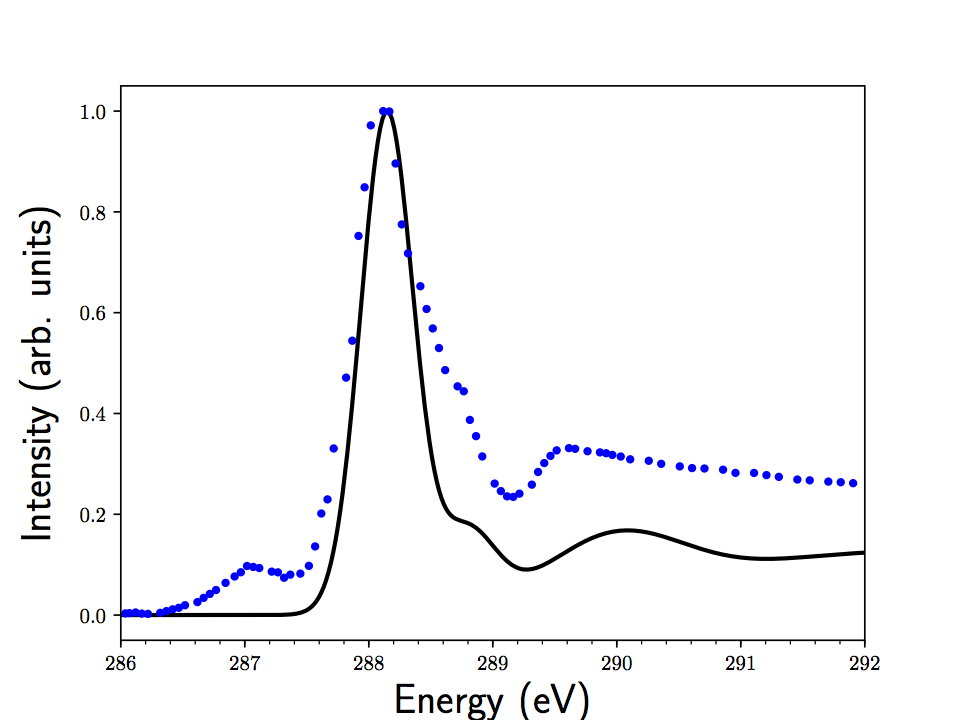}
\caption{ The X-ray spectra of methane.
The dotted (blue) line is the experimental  electron energy-loss result \citep{Hitchcock1994}. 
Full (black) line is the spectrum we calculate in this work.}
\label{fig1}
\end{figure}

\begin{figure}
\centering 
\includegraphics[scale=0.22]{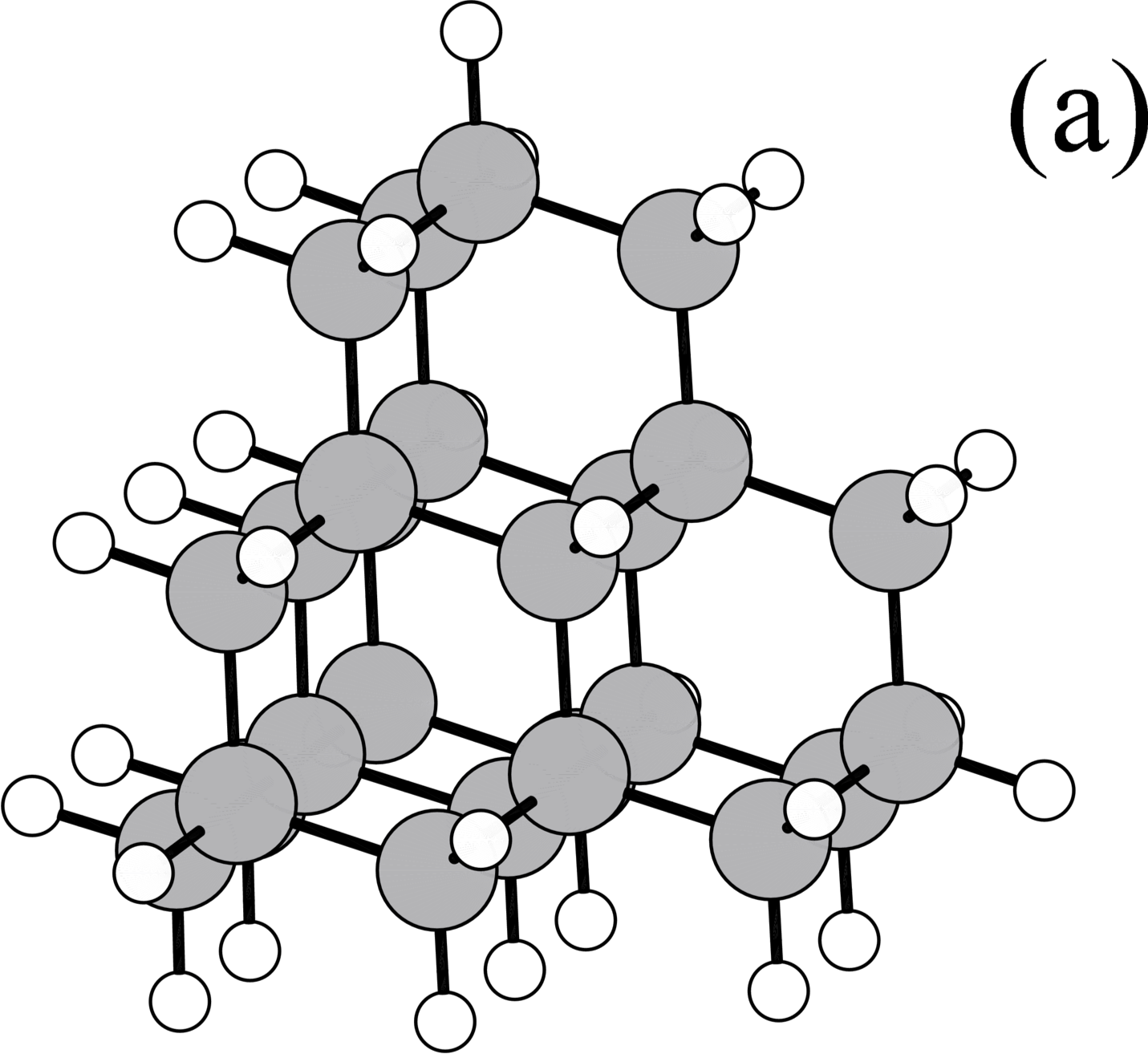}
\vskip 8pt
\includegraphics[scale=0.30]{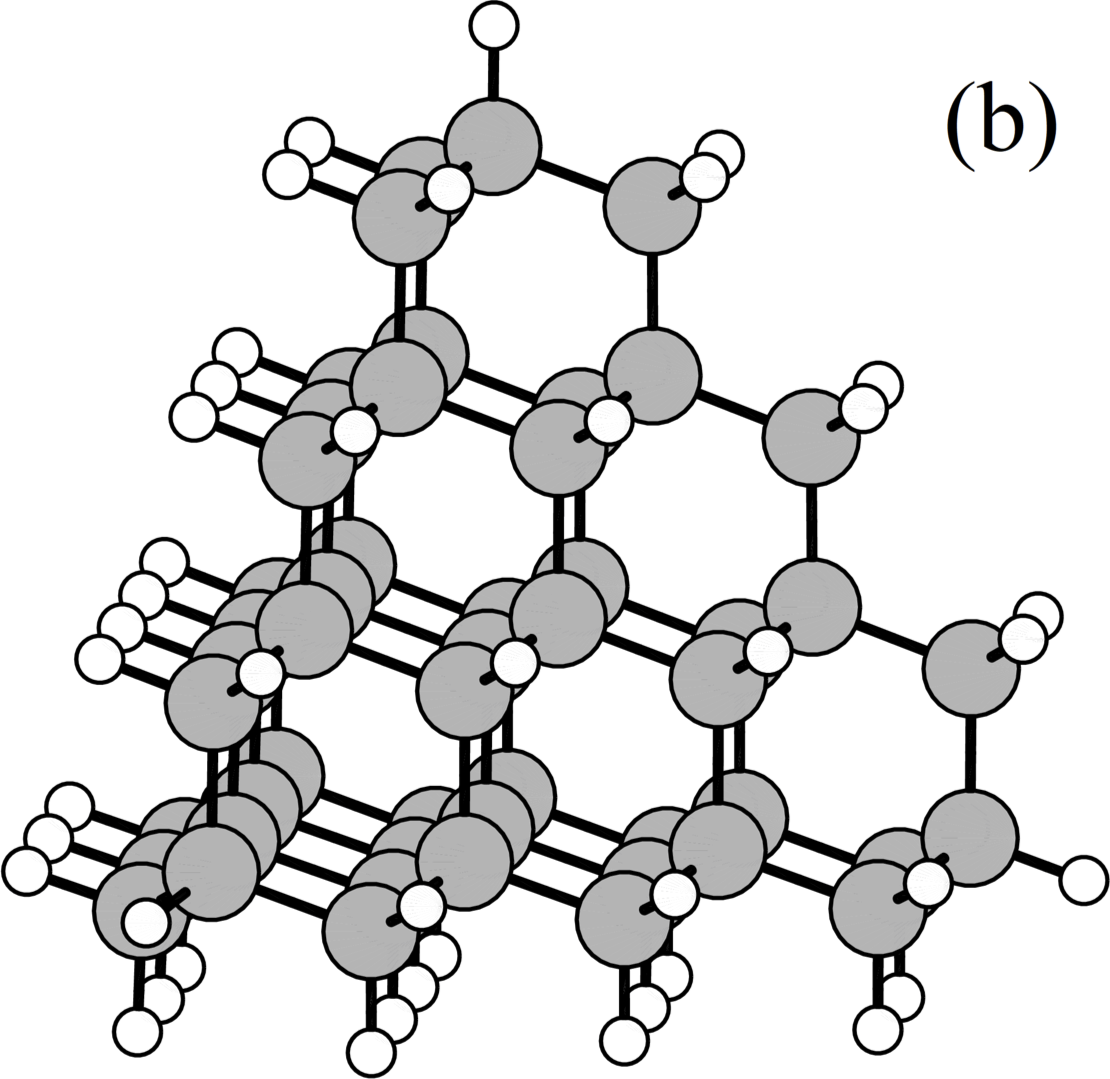}
\caption{The optimized structures of: (a) C$_{26}$H$_{32}$,
(b) C$_{51}$H$_{52}$. Darker larger balls represent carbon and lighter smaller balls hydrogen atoms.}
\label{fig2}
\end{figure}

\begin{figure}
\centering 
\includegraphics[scale=0.5]{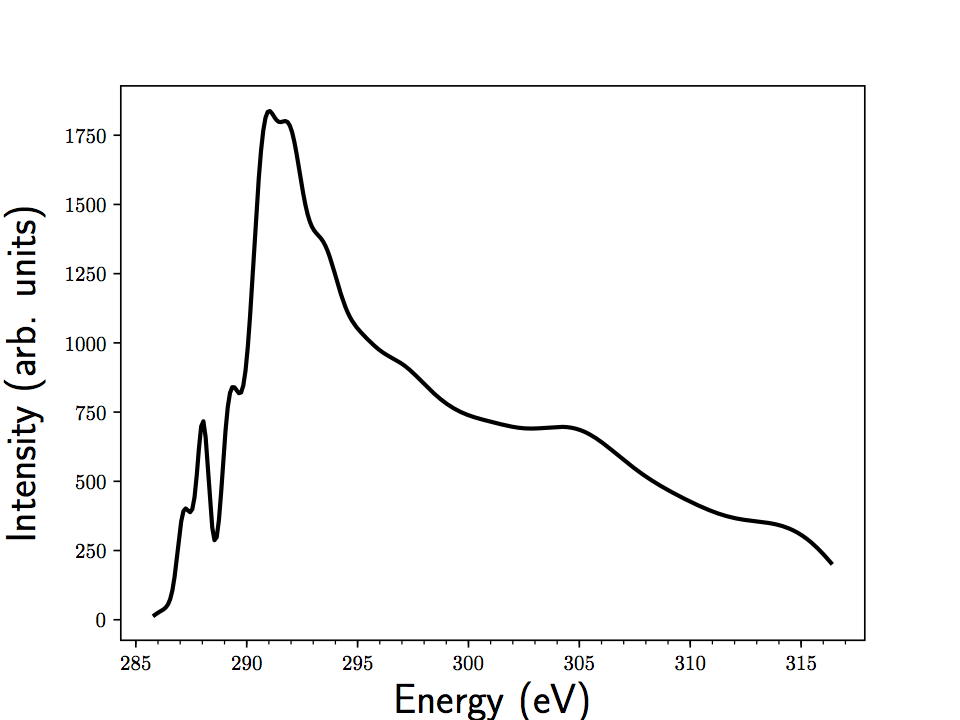}
\caption{The X-ray absorption spectrum of C$_{26}$H$_{32}$. }
\label{fig3}
\end{figure}

\begin{figure}
\centering
\includegraphics[scale=0.52]{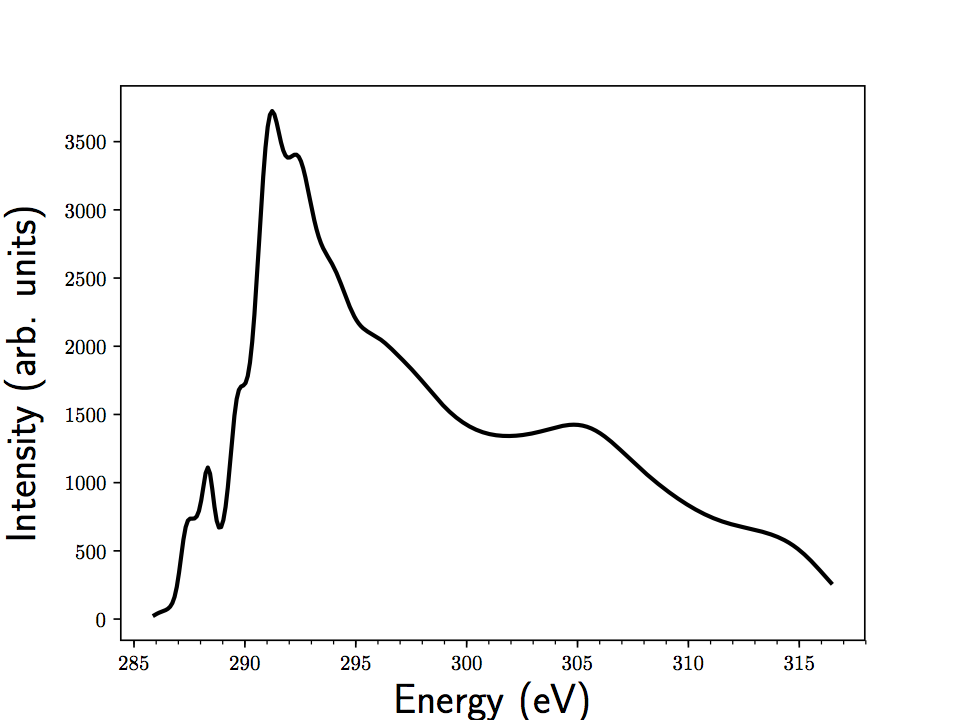} 
\caption{
The X-ray absorption spectrum of C$_{51}$H$_{52}$.}
\label{fig4}
\end{figure}

Many  carbon species exhibit resonant spectral features in the 30 eV interval above the K-edge threshold of the carbon atom.
The value of the K-edge for the carbon atom is 284 eV \citep{Elam2002}. 
Examples are X-ray absorption spectra for benzene (C$_6$H$_6$) and cyclohexane (C$_6$H$_{12}$) condensed on the substrate, as well as the bulk diamond and graphite, shown  together in Fig. 1.2 of the  book by \cite{StohrBook}. It is obvious that pre-edge features do not exist for benzene, cyclohexane, bulk diamond and graphite. However, these pre-edge features
are typical for spectra of nanodiamonds. 

\cite{Bernatowicz1990} measured the EELS spectrum of the carbonaceous meteorites Murray and Allende. Their presented spectrum for the Murray meteorite approximately resembles our result. However, pre-edge features are more pronounced in our Figs. 3 and 4. This shows that hydrogenated nanodiamonds are only one of components in the Murray meteorite. Nanodiamonds in the Allende and Murchison meteorites were analyzed by \cite{Stroud2011}. They concluded that these meteorites consist of a two-phase mixture of nanodiamonds and glassy carbon.  \cite{Stroud2011} proposed that this complex material is the product of supernova shock-waves.  The presence of glassy carbon changes the EELS spectrum of meteorites. Therefore, \cite{Stroud2011}
separately measured EELS spectra from nanodiamonds and glassy carbon regions. The spectrum of regions with
nanodiamonds is similar to the spectra shown in Figs. 3 and 4.

CO is the second most abundant molecule in the ISM. 
Its carbon K-edge is at 287.4 eV \citep{Hitchcock1994}, and it is positioned in the pre-edge region of nanodiamonds. Pre-edge features of nanodiamonds have smaller intensities in a comparison with the carbon K-edge of the CO molecule. 
In addition, the carbon K-edge of CO is linked with its oxygen K-edge at 534.2 eV 
\citep{Hitchcock1994}.  There are no features in X-ray spectra of nanodiamonds  at $>$ 500 eV. Therefore, CO does not prevent a detection of nanodiamonds from their X-ray spectra. X-ray spectra of other carbon and carbon-hydrogen species in space (for example PAHs) are not similar to the spectra of nanodiamonds. These other carbon compounds do not have a pre-edge region. X-ray spectra of nanodiamonds are specific, with its pre-edge, edge, and post-edge regions.

\begin{table}
\begin{center}
  \begin{tabular}{  | c | r }
    \hline
    Structure & a$_{max}$ (\AA)\\ 
    \hline
      C$_{26}$H$_{32}$ & 7.3 \\ 
    \hline
     C$_{51}$H$_{52}$ & 9.7 \\
    \hline     
     C$_{87}$H$_{76}$ & 11.6 \\
     \hline
      C$_{136}$H$_{104}$ & 14.7 \\  
      \hline
       bucky diamonds  & 20-30 \\  
  \end{tabular}
  \caption{DFT optimized morphologies and sizes of several diamond related carbon nanoparticles. Four large tetrahedral nanodiamonds are shown, as well as bucky diamonds. It was found  that bucky diamonds are
 more stable in the size region shown in Table 1, for a broad range of pressures and temperatures \citep{Raty2003,Galli2010}. 
 The a$_{max}$ values show the maximum of distances between atoms in a corresponding structure.} 
\end{center}
\label{table1}
\end{table}

In previous DFT calculations  it was found that   bucky diamonds are more stable than nanodiamonds for sizes (2-3) nm,
and for a broad range of temperatures and pressures
\citep{Raty2003,Galli2010}. Bucky diamonds consist of a diamond core and a fullerenelike surface layers with a mixed pentagon-hexagon network.
Hydrogen atoms are absent. Sizes of tetrahedral nanodiamonds and  bucky diamonds are compared in Table 1. It is interesting that fullerenes are also discovered in space \citep{Tielens2013}. Therefore, several diamond related nanoparticles  should be considered in astrochemistry. X-ray spectra of nanodiamonds (Figs. 3 and 4), as well as
Table 1, show that fitting of dust properties to the bulk diamond values (i.e. those measured for a diamond cubic lattice) could produce errors.  Nanodiamonds or bucky diamonds could be more stable for some conditions in space.

A recent comparison of {\it Swift} X-ray/FUV  observations with those done using
{\it ALMA} radio telescope  for the T Tauri star IM Lup pointed out the importance of X-ray flare-driven chemistry
in the circumstellar disk \citep{Cleeves2017}.  Flare-driven processes could be also present at Elias 1 and HD 97048.  A strong flare in the Elias triple system was already observed by \cite{Giardino2004}. Flares could induce the formation of nanodiamonds from amorphous and graphitic-like dust grains, as well as destroy smallest diamond nanoparticles in protoplanetary disks.
Bigger ones such as C$_{51}$H$_{52}$   should remain. This agrees with studies of IR spectra by \cite{Pirali2007}  were it was found that measured spectral intensities of large nanodiamonds correspond to astronomical observations of  Elias 1 and HD 97048.  

\subsection{Detection of nanodiamonds by X-ray telescopes}

We propose a detection of large tetrahedral nanodiamonds by future high-resolution X-ray telescopes such as {\it Arcus }
\citep{Smith2016,Brenneman2016}. {\it Arcus} is the soft X-ray grating spectrometer project submitted to NASA in 2016.
It was selected for a Phase A concept study in August 2017. {\it Arcus }
should work with an effective area $>$500 cm$^2$,  spectral resolution R=$\lambda/\Delta\lambda >$4000,
and in the (8-51) \AA\, (i.e., (0.24-1.55) keV) range.   {\it Arcus}
will study sources in the Milky Way and other galaxies.
Science questions which {\it Arcus } should answer are broad and range from baryons in the Universe, black holes, galaxy clusters and  the cosmic web, to various stars, their protoplanetary disks and exoplanets. When its baseline missions will be completed, {\it Arcus} will study additional scientific questions. 
We suggest that properties of nanodiamonds and other forms of carbon dust should be studied, as {\it Arcus} will be capable of discriminating between them, as well as between different types of diamond  nanoparticles 
(see Fig. 5).
Till now X-ray spectral studies of cosmic dust were focused on iron and silicate based 
 grains \citep{Lee2005,Lee2009,Zeegers2017,Rogantini2018}.
However, carbon astrochemistry is very rich and important to understand various processes in the evolution of galaxies, stars and planets.

For the target sources we propose to look at HAeBe stars Elias 1 (V892 Tau) and HD 97048 (CU Chamaeleontis) were the existence of nanodiamonds was confirmed by IR spectroscopy 
\citep{Guillois1999,Jones2000,VanKerckhoven2002,Habart2004,Topalovic2006,Pirali2007,Goto2009}.
These stars are also confirmed as X-ray sources by previous generations of X-ray telescopes 
\citep{Zinnecker1994,Hamaguchi2005,Giardino2004,Skinner2004,Franciosini2007}.
Additional target sources are supernovaae \citep{Tielens1987} and active galactic nuclei \citep{HaroCorzo2007,Rouan2004,Gratadour2006}.

Further, a target source need not be previously identified as rich in nanodiamonds; the diffuse ISM itself may harbor them in small quantities  such as 5\%  \citep{Tielens1987}, though possibly as high as 
10\% \citep{Lewis1989}. To examine this, we have simulated an {\it Arcus} spectrum of Cyg X-2 (see Fig. 5) using the fit of the source's Chandra/ACIS spectrum by \cite{Seward2013}. 
We modeled it as an absorbed disk blackbody, with N$_H$=1.7x10$^{21}$ cm$^{-2}$, T$_{in}$=1.5 keV, and 
F$_{X}$(0.25 - 1.25 keV)$=1.4 \times 10^{-9}$erg cm$^{-2}$ s$^{-1}$.
We assumed the proto-solar elemental abundances of  \cite{Lodders2003}. Following \cite{Kaastra1996} and \cite{Pinto2010}, we then estimated the nanodiamond cross sections by scaling them to the free atom photoionization cross sections of \cite{Verner1995}. The scaled cross sections were then merged with the free atom cross sections, so that the scaled cross sections would be used over their energy range, and the free atom cross sections would be used outside that range.
\cite{Leger1991} found that PAHs account for 10\% of the carbon in the ISM. To approximate the PAHs contribution, we obtained the experimental EELS of benzene (C$_6$H$_6$) from the online database \citep{Hitchcock1994} and scaled it in the same way as the nanodiamond spectra. 
We then made a series of simulated spectra, setting quantities of C$_6$H$_6$ and carbon gas to 10\% and 80\% of total available carbon, respectively, with the remaining 10\% in nanodiamonds. We fitted the spectra from (280-330) eV and allowed all parameters to float. An implementation of the Cash statistic, cstat, was used to find the best fits \citep{Cash1976,ArnaudBook}. When the nanodiamond contribution was set to be either entirely C$_{26}$H$_{32}$ or C$_{51}$H$_{52}$, they could be detected to 5 $\sigma$ in 200 ks. Next, we tested to see if we could distinguish between diamond types by
fitting the spectrum with all C$_{26}$H$_{32}$ absorption with a model using C$_{51}$H$_{52}$. This produced a goodness-of-fit value (cstat/degrees of freedom) of 2660/2607. For comparison, the value obtained from fitting it with a model using C$_{26}$H$_{32}$ was 2643/2607, indicating that the nanodiamond types can be distinguished. Similar results were obtained when testing the spectrum with all C$_{51}$H$_{52}$ absorption in the same way. The simulated spectra with all C$_{26}$H$_{32}$ and all C$_{51}$H$_{52}$ nanodiamonds are shown in the top and middle panels of Fig.5, respectively. We also considered the case where C$_{26}$H$_{32}$ and C$_{51}$H$_{52}$ each accounted for 5\% of the total C available. The exposure time required for 5 $\sigma$ detections of both diamond types rose to 4 Ms (see Fig.5, bottom panel). While is this a large time investment, determining the nanodiamond size distribution in the diffuse ISM could shed light on diamond formation and destruction \citep{Tielens1987}. Further, both the oxygen K and iron L edges fall within the {\it Arcus} bandpass, and such an observation would allow an examination of silicate and iron-bearing dust features in unprecedented detail at those edges, as well.

\begin{figure}
\centering
\includegraphics[scale=0.6]{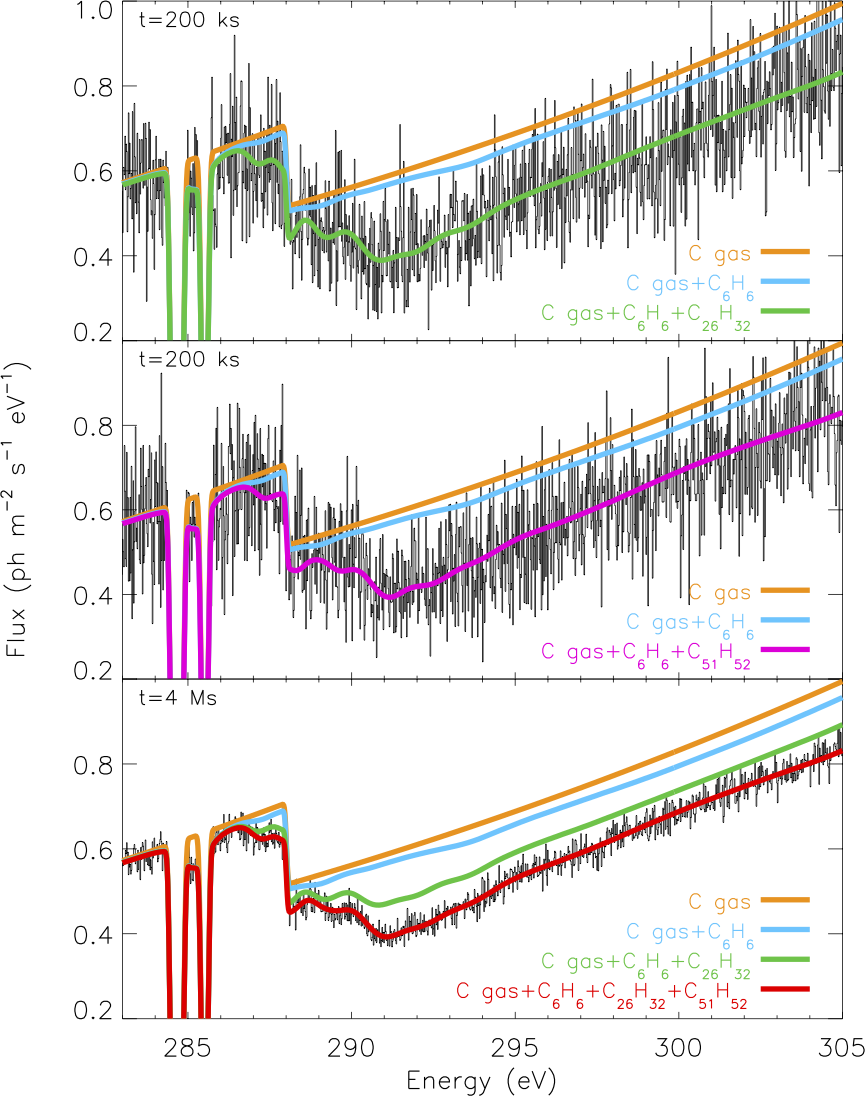} 
\caption{
Simulated {\it Arcus} spectra of Cyg X-2 at the carbon K edge, for different exposure times, as well as
nanodiamond types and amounts.}
\label{fig5}
\end{figure}

\section{Conclusions}
\label{concl}

Nanodiamonds have been a subject of research in astrophysics for a long time. They have been involved in diverse topics such as Earth and planetary astrophysics, the interstellar medium, stars, supernovae, pulsars, and active galactic nuclei.  However, nanodiamonds are still rather unexplored type of cosmic dust.  We calculate 
the X-ray absorption spectra at the carbon K-edge of tetrahedral hydrocarbon nanoparticles CH$_4$, C$_{26}$H$_{32}$, and
C$_{51}$H$_{52}$  using density functional  theory methods. A specific combination of pre-edge, edge, and post-edge
features differs from  spectra of other carbon and C-H species and could be used for detection of nanodiamonds by X-ray spectroscopy. We suggest detection studies of nanodiamonds (as well as other forms of carbon based cosmic dust) 
by future X-ray telescopes, such as the project {\it Arcus} recently selected by NASA for a Phase A concept study.
In particular, active galactic nuclei are favorable environments for the formation of nanodiamonds. Diamond nanoparticles  efficiently form in UV shocks and they could survive harsh environments where silicate dust is destroyed. Strong target candidates to detect nanodiamonds by X-ray spectroscopy are X-ray active Herbig Ae/Be stars Elias 1 (V892 Tau) and HD 97048 (CU Chamaeleontis).

The results of DFT calculations can be scaled and used in ISM spectral fitting. Here, to simulated {\it Arcus} spectra we have applied the method that was used to scale similar molecular spectra (i.e., spectra in arbitrary flux units).
Using the scaled spectra, we have simulated an observation of absorption in the diffuse ISM 
(N{$_H$}=10$^{21}$cm$^{-2}$) toward a bright source (Cyg X-2), using standard values for ISM carbon abundance. We have set the amounts of carbon in diamonds and PAHs each to 10\%, respectively, as suggested by the literature; when only one type of diamond was used, the absorption from nanodiamonds could be detected to 5 $\sigma$ and the diamond type could be distinguished in a 200 ks observation.

\section*{Acknowledgements}
Calculations presented in this work were
done using the computational cluster Isabella at the University of Zagreb Computing Centre SRCE.  GB acknowledges the support of the Croatian Science Foundation (HRZZ) grant IP-2014-09-8656. 
We would like to thank Elisa Costantini and Sascha Zeegers for discussions.
We thank the referee for useful suggestions.
This research has made use of NASA's Astrophysics Data System Bibliographic Services.

\bibliographystyle{mn2e} 
\bibliography{xd}
\label{lastpage}

\end{document}